\documentclass[twocolumn,letterpaper,prb]{revtex4}
\usepackage{graphicx}
\usepackage{dcolumn}
\usepackage{amsmath}
\begin{document}

\title{ Evaporative Cooling of Helium Nanodroplets with Angular
Momentum Conservation}

\author{ Kevin K. Lehmann}
\email[]{ Lehmann@princeton.edu}
\author{ Adriaan M. Dokter}
\affiliation{ Department of Chemistry, Princeton University,
Princeton NJ 08544}

\date{\today}

\begin{abstract}
Evaporative cooling of helium nanodroplets is studied with a
statistical rate model that includes, for the first time, angular momentum conservation
as a constraint on the accessible droplet states. It is found that
while the final temperature of the droplets is almost identical to
that previously predicted and later observed, the distribution of
total droplet energy and angular momentum states is vastly more
excited than a canonical distribution at the same temperature. It
is found that the final angular momentum of the droplets is highly
correlated with the initial direction, and that a significant
fraction of the alignment of the total angular momentum should be
transferred to the rotational angular momentum of an embedded
molecule.
\end{abstract}

\pacs{}

\maketitle

The study of helium nanodroplets has received considerable attention
in the past decade.~\cite{Goyal92,Toennies98,JCP_special_issue}
Such droplets rapidly cool by helium atom
evaporation while they travel inside a vacuum chamber.
Brink and Stringari~\cite{Brink90} used a statistical evaporation model
and predicted a terminal
temperature of $^4$He nanodroplets of ~0.4\,K, in excellent agreement
with the value of 0.38\,K later deduced from the rotational structure
in vibrational spectra of SF$_6$ and other molecules.~\cite{Hartmann95,Callegari01}
Despite the obvious success of this theoretical work, the model used is clearly
incomplete in that the constraint of angular momentum conservation
was not imposed.  The need for a more complete evaporative cooling
study was made evident by the recent observation of a polarization
anisotropy in the absorption spectrum of pentacene in helium droplets.~\cite{Portner03}
The authors of this work suggested that the total angular momentum deposited
in the droplets by the pickup of a heavy molecule is aligned perpendicular to
the droplet velocity, and that this droplet alignment survives the evaporative
cooling and is transferred to the embedded molecule.
The present study was undertaken to test the reasonableness of this
conjecture.

We model the evaporative cooling with a statistical rate approach,
analogous to phase space theory in unimolecular dissociation,
which explicitly includes the constraints of angular momentum
conservation.~\cite{Baer96} We use Monte Carlo sampling to follow
cooling `trajectories' as the droplets lose much of their
initial energy and angular momentum by helium atom evaporation. It
is found that the droplets cool to final temperatures close to
those predicted without angular momentum
constraints.~\cite{Brink90} However,
the distribution of terminal droplet states (where the evaporative
cooling lifetime becomes longer than typical flight times in
experiments) cover a vastly broader range of energy ($E$) and
total angular momentum ($J$) than was previously expected.
Further, it is found that the final angular momentum vector of the
droplet is highly correlated with the initial value, such that
much of the alignment remains, and that a sizable fraction of this alignment
is transfered to an embedded rotor.

\textit{ Evaporative Cooling Model}---
Consider a helium nanodroplet, D, with initial values of the conserved
quantities $n$ (number of helium atoms), $E'$ (total internal energy
in units of Kelvin),
and $J'$ (the total angular momentum in units of $\hbar$).
If $E'$ is sufficiently great, the droplet will cool by helium atom
evaporation by the reaction: 
$\text{D}(n,E',J') \rightarrow \text{D}(n-1,E'',J'') + \text{He}\left(E_{\text{trans}},L \right)$
where $E_{\text{trans}}$ is the center of mass translational
energy in the dissociation and $L$ is the orbital angular momentum
quantum number of the fragments. Conservation of total angular
momentum requires that $J', J'', L$ obey the triangle rule.
Based upon the bulk density of helium,~\cite{Donnelly98}
the droplet has radius $R(n) = r_0 n^{1/3}$ with $r_0 =
2.22\,$\AA. Conservation of energy, including the change in the
surface energy defined by surface tension, $\sigma =
0.272$\,K\AA$^{-2}$,~\cite{Deville96} requires that
\begin{equation}
E' = E'' + E_{\text{b}} - \frac{8\pi}{3} r_0^2 \sigma n^{-1/3} + E_{\text{trans}}
\end{equation}
where $E_{\text{b}} = 7.2\,$K is the binding energy of helium to the bulk.~\cite{Donnelly98}
Because of the centrifugal barrier, classically
$E_{\text{trans}} \ge  \hbar^2 L(L+1)/\left(2 m_{He} R(n)^2 \right)$,
leading to a $E''_{\text{max}}$ value for each $L$.

Using statistical reaction rate theory, the rate of helium atom evaporation
is given by~\cite{Baer96}
\begin{equation}
k(n,E',J') = \frac{1}{h} \frac{N_{\text{o}}(n,E',J')}{\rho(n,E',J')}
\end{equation}
$\rho(n,E',J')$ is the density of states of the droplet at
fixed angular momentum, and
$N_{\text{o}}(n,E',J')$
equals the number of states of the product droplet and departing
He atom consistent with total $E'$ and $J'$ and with the 
departing He atom having kinetic energy above the centrifugal barrier.
If we denote the total number of
$n$ atom droplet states with energy $\le E$ and
total angular momentum quantum number $J$ by $N(n,E,J)$, we can use
the triangle rule to write $N_{\text{o}}(n,E',J')$ as a sum of 
$N(n-1,E''_{max}(L),J'' )$ over all allowed values of $L, J''$.

For droplets of the size and energy range of interest to
experiments, the density of states is dominated by
quantized capillary waves on the surface of the droplets 
( ripplon modes).~\cite{Brink90} 
Simple, but highly
accurate, analytical approximations for $N(n,E',J')$ and
$\rho(n,E',J') = \left(
\partial N/ \partial E \right)_{J'}$ have recently been published,~\cite{Lehmann03a}
 thus allowing
calculation of $k(n,E,J)$. 
Using an ensemble that conserves $n, E, J$, the droplet temperature can be
calculated using $(k_{\text{B}} T(n,E,J))^{-1}
 = \left( \partial \ln \rho(n,E,J)/ \partial E \right)_{J}$.
For droplets with only ripplon excitations, the densities of
states depends upon $n$ only through the size of a reduced energy
scale which is equal to $3.77 n^{-1/2}\,$K. Starting with a given
initial size, energy, and total angular momentum for the droplet
at time zero, we calculate the evaporation rate and advance time
by the inverse of this number. We then randomly pick a single open
decay channel ($E'',J'',L$) with probability proportional to
$N(n-1,E'',J'')$.
We can treat the product droplet as a new
initial condition, and calculate another evaporation event.  As
the droplet cools, the rate of evaporation falls exponentially.
When the cumulative time for the next evaporation event is greater
than the assumed free flight time in an experiment, we terminate
the evaporation process.  

We also did evaporative cooling calculations for droplets with an
embedded linear rotor with effective rotational constant $B$,
which will typically be several times smaller than the
corresponding value in the gas phase due to helium motion induced
by the rotation.~\cite{Hartmann95,Callegari01} For these
calculations, the integrated density of states,
$N_c(n,E,J)$, for the combined droplet + rotor system is
calculated by convolution of the integrated droplet density of
states with the spectrum of the rigid rotor:
$N_c(n,E,J) = \sum_{j} \sum_{J' = |J - j|}^{J + j} N(n,E - B
j(j+1), J')$
These were calculated for various values of $B$ and the resulting
densities fit to the same functional form previously used for
$N(n,E,L)$, but with the fitting constants now expressed as
polynomial function of $B$ and its inverse.  Differentiation gives
the combined density of states, $\rho_c(n,E,J)$.

Alignment is defined as $\left<P_2(\cos\theta)\right>$, where $\theta$ is the
angle of the angular momentum vector with the quantization axis
(chosen to be the velocity vector of the helium droplet beam).
In order to track the changes in alignment during cooling,
we must propagate the angular momentum
projection quantum numbers of droplet ($M$) and embedded rotor ($m$). 
The final alignments are found to be proportional to the initial alignment of
the total angular momentum created by the pickup process.

If we have an initial 
distribution $P(n,M')$ over angular momentum projection quantum
numbers $M'$, we can use the Wigner-Eckart theorem to calculate the
product distribution of the projection quantum number
populations:~\cite{Edmonds}{
$P(n-1,M'') = \sum_{M'}  \left< J'', M'', L, M' - M'' | J', M'
\right>^2 P(n,M')$.
The assumption of equal probability of all
states consistent with total $E$ and $J$ leads to
a similar expression for the probability of
populating a rotor quantum state $j,m$.
This allows the calculation of the alignment of the rotor
angular momentum, which is an experimental observable.

\textit{Results}---
We did evaporative cooling calculations with initial values of $n
= 10^4$, $E = 1700.$\,K, and $J = 10, 1000, 2000, \dots 5000$,
computing 2500 Monte Carlo `trajectories' in each case.
These values where selected as likely initial conditions following
the pickup of a tetracene molecule.~\cite{Portner03,Lehmann03b}
These conditions correspond to an initial droplet temperature of
$2.61\,$K independent of $J$. We also assumed that the initial condition for the
total angular momentum is $P(M) = \delta_{0,M}$ (initial alignment $= -1/2$). 
A cooling time of $100\,\mu$s was used.

We can first make some qualitative predictions.   We expect that on average
the evaporated helium atoms will reduce droplet energy by $E_{\text{b}} + (3/2)k_{\text{B}}T$ where
$T$ is the average temperature during evaporation, which would imply 
an energy loss of $\approx 10$\,K per He evaporated.   
Thus, we would expect $\approx 170$ helium atoms to be evaporated.
A helium atom evaporated at  $T$ will carry root mean squared (rms)
angular momentum of  $\sqrt{2 m_{He} k_B T} R \approx 23 \hbar$.  If the 
departing atom's
angular momentum is parallel to the initial angular momentum
(as for water droplets leaving a wet, spinning tennis ball), we can expect the
droplet to lose about $4000 \hbar$ while cooling.
Alternatively, if the orbital angular momenta of the evaporated atoms are
random, the droplet angular momentum would undergo a 3D random walk, and the mean squared
angular momentum of the droplets would be expected to \textit{increase} by the number
of evaporated atoms times the square of the angular momentum per evaporation event.
The form for $\rho(n,E,J)$ for fixed $n,E$ is that of a thermal distribution
of a spherical rotor with a mean value of $J(J+1)$ that grows as $E^{8/7}$.~\cite{Lehmann03a}
For the initial values of $n,E$, the rms of the distribution of $\rho(n,E,J)$
occurs for $J = 345$, an order of magnitude below the mean value of
the angular momentum deposited in pickup of a large molecule, such as
pentacene.  It can be expected that during evaporation, the density of states will
bias evaporation such that the
$J$ moves towards the rms value, which will result in the average angular
momentum being reduced during evaporation.  
For a canonical distribution at $0.38\,$K, the mean terminal value of $\left<E\right> = 18.5$\,K
and $\sqrt{\left<J(J+1)\right>} = 39$ are predicted for droplets of $n =10^4$.~\cite{Lehmann03a}

Figure~\ref{fig:trajectory} shows a plot of droplet energy, angular momentum
quantum number, temperature, and number of evaporated helium atoms
 as a function of time for five representative
trajectories with initial condition $E_{int} = 1700.$\,K and
$J_{int} = 3000$.   It is found that most of the evaporation takes
place in the first $\approx 1$\,ns, but that the temperature
slowly drops even for long times. The final temperatures of each
trajectory is similar to that observed experimentally for
molecules inside helium droplets~\cite{Hartmann95,Callegari01} and
also to that predicted by evaporative cooling calculations without
angular momentum constraints.~\cite{Brink90} The remarkable
difference from those previous calculations is that the final
energies and angular momenta are vastly higher than that found in
the previous work.

\begin{figure}[htbp]
\includegraphics[width=3in]{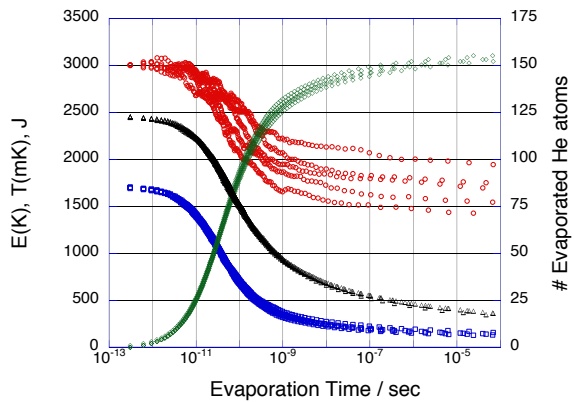}
  \caption{Five evaporative cooling trajectories starting with initial conditions $J=3000\,\hbar$, $E = 1700.\,$K,
$n = 10^4$, which predicts an initial droplet temperature of $\approx 2.5\,$K.   Plotted are the droplet angular
momentum, temperature, energy, and number of helium atoms evaporated as a function of time since the start of
the simulation}
  \label{fig:trajectory}
\end{figure}

Table I gives a summary of the numerical results for a
range of initial angular momentum values.  
For calculations with a shorter cooling time ($10\,\mu$s), the results
were almost identical, except that the final temperatures were
systematically $\approx 50$\,mK higher.
It is seen that droplet
cooling gives a mean residual energy and angular momentum vastly
higher than that of a canonical ensemble at the same final
temperature, and this energy and angular momentum rises quickly
with initial angular momentum value.   Notice that the `trapping'
of angular momentum is found even for initial values considerably
smaller than the `maximum' value predicted above that can be shed
by evaporation.  It is also evident from the table that most of
the initial alignment of the total angular momentum is maintained
which indicates that the final angular momentum is nearly parallel
to the initial angular momentum.  While for each initial
condition, a broad distribution of final $E$ and $L$ values are
found, these are distributed in a narrow band of energy width $\approx 8$\,K,
following a line corresponding to constant temperature. 
For fixed $J$, the final $E$ values are $\approx 45$\% larger than the
lowest possible droplet energy for that $J$, which corresponds to
$J/2$ quanta in the lowest, $L =2$ ripplon mode.

\begin{table*}[htb]
\label{tab1}
\centering
\caption{Evaporative Cooling of Helium Nanodroplets initially with 10000 helium
atoms and initial internal energy equal to $1700.$\,K.  ``$\pm$'' after each entry
gives the standard deviation of the ensemble of 2500 cooling trajectories.  }
\begin{tabular}{lccccccr}
\hline\hline
$J \text{initial}$       & $10$              &   $1000$          &      $2000$        & $3000$           &  $4000$ & $5000$&  \\
\hline
$\left< \Delta n\right>$             &$166.05 \pm 3.15$ &$164.39 \pm 3.09$  & $160.75 \pm 2.86$  & $157.24 \pm 2.83$& $153.63 \pm
2.68$       &$150.05 \pm 2.46$    &  \\
$ \left<E_{\text{final}}\right>$     &  $25.88 \pm5.66$  &$44.45 \pm 8.56$   &  $81.06 \pm 9.96$  &  $125.0 \pm 10.1$& $174.98 \pm
10.30$        & $229.2 \pm 9.9$ & K  \\
$ \left<J_{\text{final}}\right>  $   & $170. \pm 64.$    &$404 \pm 107.$     &  $897. \pm 135.$   &  $1528. \pm 145.$& $2273. \pm 152.$
& $3108. \pm 151.$ &  \\
$ \left<T_{\text{final}}\right>  $   & $354.7\pm 21.9$   &$347.1 \pm 17.2$   & $343.4 \pm 11.4$   &  $342.8 \pm 8.2$ & $344.4 \pm 6.2$
& $346.1 \pm 5.0$      &  mK  \\
$\left<E_{\text{trans}}\right>$                  & $3.53 \pm 2.79$   &$3.52 \pm 2.75$    & $3.53 \pm 2.71$    &  $3.47 \pm 2.65$ & $3.39 \pm 2.56$
&$3.26 \pm 2.44$     &K \\
$\sqrt{\left<L(L+1)\right>}$        & $25.6 \pm 26.2$   &$25.7 \pm 26.0$    & $26.1 \pm 26.1$    & $26.5 \pm 26.1$  & $26.6 \pm
25.9$        &     $26.6 \pm 25.6 $ & \\
$\left<\cos(\theta_{J',L})\right>$  & $0.03 \pm 0.58$   &$0.156 \pm 0.572$  & $ 0.267 \pm 0.554$ & $0.357 \pm 0.530$& $0.426 \pm
0.504$       &$0.480 \pm 0.480$    \\
$\frac{\left<J\text{(final) alignment}\right>}{J\text{(initial) alignment}}$
                    & $ 0.002 \pm 0.020$     &$0.829 \pm 0.053$  &$0.958 \pm 0.007$   &$0.983 \pm 0.002 $& $0.992 \pm 0.001$
&$0.995 \pm 0.0004 $ &  \\
 \hline\hline
  \end{tabular}
\end{table*}

While a systematic study of the final distributions upon the full range of
likely droplet initial conditions is beyond the scope of the present report,
we would like to indicate the trends.    If the initial energy and total
angular momentum quantum number are kept constant at $1700.$\,K and $1000$ respectively,
and the initial size of the droplet is decreased from $n = 10^4$ to  $3 \times10^3$,
the initial temperature rises from 2.61 to 3.69\,K.   As expected,
the higher initial temperature rises the average kinetic energy carried away
by the He atoms ($3.53 \rightarrow 4.88$\,K), resulting in a modest decrease
($164 \rightarrow 149$) in the average number of evaporated He atoms. Because of decreased
droplet size, the rms orbital angular momentum carried away by the He atoms decreases
($25.7 \rightarrow 20.4$).  Despite this, the final average angular momentum decreases
($401 \rightarrow 355$), due to an increased correlation in the emission direction
(average cosine of angle between $L$ and $J$ increases from 0.156 to 0.23).
The final average energies and temperatures are almost unchanged.
If we keep the initial angular momentum at $J = 1000$ and the size at $n = 10^4$
but decrease the initial energy from $E = 1700.$ to  $1000.$\,K,
we find that the initial temperature ($2.61 \rightarrow 2.08$\,K),
and number of evaporated helium atoms ($164 \rightarrow 101$) both decrease
as expected.   The He atoms mean kinetic energy ($3.53 \rightarrow 2.87$\,K)
and rms orbital angular momentum ($25.7 \rightarrow 23.0$) both decrease
as expected for lower initial temperature.  Unexpectedly, the average
final droplet energy ($44.2 \rightarrow 49.8$) and angular momentum
($401 \rightarrow 475$) increase when the initial energy is decreased.

We performed evaporative cooling calculations for $B = 0.6$ and 1.2\,GHz.
These values were selected as they represent typical values for ``heavy rotors'' that
have been studied in helium nanodroplets with rotational resolution.~\cite{Callegari01}
As expected, the presence of the rotor has little effect upon the final distribution
of $E$ and $J$ of the droplets. 
Figure~\ref{fig:rotor_alignment} shows the calculated ratio of alignment of the
rotor angular momentum to that of the total angular momentum for droplets with
the average final values of $E$ and $J$ for initial states corresponding to the
$J_{\text{initial}} = 10, 1000, 2000 \ldots 5000$.  It can be seen that
the degree of rotor alignment increases with both $j$ and the initial angular
momentum, though the alignment ratio appears to saturate at higher $J_{\text{initial}}$
values.  The level of alignment found in this work is certainly within the
range that should be detectable by experiments of the type reported for
pentacene ($\approx 10-20$\%).~\cite{Portner03}

\begin{figure}[htbp]
\includegraphics[width=3in]{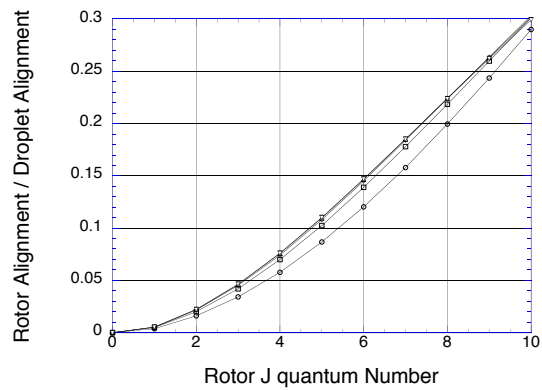}
  \caption{Plot of the ratio of the alignment of an embedded rotor to the alignment
of the total angular momentum of the droplet, as a function of the rotational quantum number of the rotor.
Curves are for final states reached from initial conditions $E = 1700.\,$K and $J = 1000, \ldots 5000$
starting from the lowest curve}
  \label{fig:rotor_alignment}
\end{figure}

\textit{Experimental Consequences}--
This work has demonstrated that the distribution of the internal
excitations of helium nanodroplets should be vastly more excited
than had been previously predicted, based solely upon the low temperature
of the droplets.   The present results support the interpretation of
P\"ortner \textit{et al.},
that a remnant of the initial pick up angular momentum survives
evaporative cooling and provides a bias that partially aligns
embedded molecules.~\cite{Portner03}  The present work predicts
that this should be a common phenomenon. It also provides a
rationalization for the failure of a previous attempt to predict
the rovibrational lineshapes of embedded molecules based upon the
inhomogeneous distribution of molecule ``particle in a box''
states.~\cite{Lehmann99b}  That work assumed that these
translational states followed a Boltzmann distribution, while the
present work suggests that high angular momentum translational
states of the embedded molecules will be substantially more
populated in the droplets.   More significantly, the dramatic
effect of angular momentum constraints calls into question the use
of Path Integral Monte Carlo simulations for nanodroplet
experiments since such calculations assume a canonical ensemble.
The present work demonstrates droplets prepared in existing
experiments are poorly described as a canonical ensemble.

Many of the predictions of the present calculations could be directly experimentally tested.
The trapping of increasing energy in the droplets with increasing initial angular momentum will reduce the expected amount of helium atom evaporation as the collision energy is changed.  However, it is not clear if this could be disentangled from other effects, including a collision and
impact parameter dependence of the pickup probability and the possibility of coherent
ejection of helium atoms during the pickup process, i.e. the early evaporation
events could be strongly nonstatistical and thus poorly predicted by the present model.
Measurement of the spatial distribution of
helium atoms evaporated from a droplet would be very revealing, as the present calculations
predict that the atomic velocity distribution is highly anisotropic in the center of mass frame.
Since the orbital angular momentum of the evaporated atoms is on average parallel to the initial angular momentum,
which itself is largely perpendicular to the droplet velocity, one expects to find the
evaporated helium atoms peaked ahead and behind the droplets in the center of mass frame.  Measurement of the
radial distribution of atoms or molecules embedded in helium nanodroplets would also
reveal the expected increased translational angular momentum of the dopant.   It is possible
that high energy electron or X-Ray scattering experiments could reveal the radial
distribution function of atoms or molecules solvated by helium nanodroplets.
The most direct test of the present model would be to measure the trapped internal energy and/or
angular momentum of the droplets.   One way that could be done in principle is to measure the Stokes and antistokes Raman spectrum from the lowest, $L = 2$ ripplon mode.   
It should be noted that excitation of the $L = 2$ mode should dominate in the trapped states, which have nearly the lowest possible internal energy for a given total angular momentum.

This work was supported by the National Science Foundation



\end{document}